\documentclass[twocolumn]{aastex61}

\usepackage{amsmath, amsthm, amssymb}
\usepackage{arydshln}
\usepackage{bm}
\usepackage{xcolor}                                                                       
\usepackage{colortbl} 
\received{July 1, 2016}
\revised{September 27, 2016}
\accepted{\today}
\submitjournal{ApJ}

\shorttitle{Fragmentation of self-gravitating disks}
\shortauthors{Deng et al.}

\begin{document}

\title{Convergence of the critical cooling rate for protoplanetary disk fragmentation achieved; the key role
of numerical dissipation of angular momentum } 

\correspondingauthor{Hongping Deng}
\email{hpdeng@physik.uzh.ch}

\author{Hongping Deng}
\affiliation{Center for Theoretical Astrophysics and Cosmology, Institute for Computational Science, University of Zurich, Winterthurerstrasse 190, 8057 Zurich, Switzerland}

\author{Lucio Mayer}
\affiliation{Center for Theoretical Astrophysics and Cosmology, Institute for Computational Science, University of Zurich, Winterthurerstrasse 190, 8057 Zurich, Switzerland}

\author{Farzana Meru}
\affiliation{Institute of Astronomy, University of Cambridge, Madingley Road, Cambridge, CB3 0HA, UK}
\begin{abstract}
We carry out simulations of gravitationally unstable disks using smoothed particle hydrodynamics(SPH) and 
the novel Lagrangian meshless finite mass (MFM) scheme in the GIZMO code 
\citep{Hopkins2015}. Our aim is 
to understand the cause of the non-convergence of the cooling
boundary for fragmentation reported in
the literature.
We run SPH simulations with two different artificial viscosity implementations, and compare them with MFM,
which does not employ any artificial viscosity.
With MFM we demonstrate convergence of the critical cooling time scale for fragmentation at $\beta_{crit} \approx 3.$.
Non-convergence persists in SPH codes, although it is
significantly mitigated with schemes having reduced
artificial viscosity such as inviscid SPH (ISPH) \citep{Cullen2010}.
We show how the non-convergence problem is caused by 
artificial fragmentation triggered by excessive 
dissipation of angular momentum in domains  with large velocity derivatives. With increased 
resolution such domains become more prominent. Vorticity lags behind 
density due to numerical viscous dissipation in these regions, promoting
collapse with longer cooling times. Such effect is shown to be dominant over the competing tendency of artificial viscosity to diminish with increasing resolution. 
When the initial conditions are first relaxed for several orbits, the flow is more
regular, with lower shear and vorticity in non-axisymmetric regions, aiding convergence.
Yet MFM is the only method  that converges exactly.
Our findings are of general interest as numerical dissipation via  artificial viscosity or advection errors can also occur in  grid-based codes. Indeed for
the FARGO code values of  $\beta_{crit}$ significantly
higher than our converged estimate
have been reported in the literature.
Finally, we discuss implications for giant planet 
formation via disk instability.

\end{abstract}

%% Keywords should appear after the \end{abstract} command. 
%% See the online documentation for the full list of available subject
%% keywords and the rules for their use.
\keywords{Gravitational instability ---accretion, accretion disks --- fragmentation ---numerical viscosity}

%% From the front matter, we move on to the body of the paper.
%% Sections are demarcated by \section and \subsection, respectively.
%% Observe the use of the LaTeX \label
%% command after the \subsection to give a symbolic KEY to the
%% subsection for cross-referencing in a \ref command.
%% You can use LaTeX's \ref and \label commands to keep track of
%% cross-references to sections, equations, tables, and figures.
%% That way, if you change the order of any elements, LaTeX will
%% automatically renumber them.

%% We recommend that authors also use the natbib \citep
%% and \citet commands to identify citations.  The citations are
%% tied to the reference list via symbolic KEYs. The KEY corresponds
%% to the KEY in the \bibitem in the reference list below. 

\section{Introduction} \label{sec:intro}
Massive rotating gas disks can become unstable due to their own self-gravity. Using linear perturbation theory 
applied to an axisymmetric, razor-thin disk,  
\citet{Toomre1964} showed that the exponential growth of 
local density perturbations is controlled by the parameter

\begin{equation}
  Q\equiv \frac{c_{s}\kappa}{\pi G \Sigma} 
\end{equation}

 where $c_{s}$ is the sound speed in the disk, $\kappa$ is the epicyclic frequency, which equals the angular frequency $\Omega$ for a Keplerian disk, and $\Sigma$ is the mass surface density ($G$ is the gravitational constant).
When $Q < Q_{crit}\approx 1$,  gravitational instability occurs locally,
and a gas patch will undergo collapse as gravity
overcomes both pressure gradients and shear induced by
rotation.

In realistic finite-thickness disks subject to generic
perturbations the character of the instability is global,
whereby the disk first undergoes a phase of
non-axisymmetric instability, namely develops a prominent
spiral pattern, before becoming locally unstable to collapse \citep{Durisen2007}. Inside and across spiral arms 
fluid elements are subject to torques that lead to
angular momentum transfer \citep{Cossins2009}. Angular momentum transfer is
indeed a key feature of gravitational instability,
and ultimately determines the evolution of the density
field in other over-dense regions inside spiral arms \citep{Mayer2004, Boley2010}.
The other central aspect is that the spiral pattern causes
shocks in the flow, releasing gravitational energy into heat
,  and eventually raising $Q$
above the stability threshold. More specifically, whether
or not $Q$ will remain low enough for the instability
to grow, and eventually lead to fragmentation, ultimately
depends on the ability of the gas to release the heat
via radiative cooling.
\citet{Gammie2001} proposed to parameterize the local cooling time scale $t_{cool}$ as
\begin{equation}
\beta=t_{cool}\Omega
\end{equation}
where
\begin{equation}
t_{cool}=u(\frac{du}{dt})^{-1}
\end{equation}
$u$ is the specific internal energy. He showed that when $\beta<\beta_{crit}\approx3$ fragmentation occurs in local shearing sheet simulations with a ratio of specific heats $\gamma=2$. Physically this means cooling has to occur
on a timescale comparable to the local orbital time, which
can be easily understood as the spiral pattern that generates shock
heating also evolves on the local orbital time.
Using 3D SPH simulations, \citet{Rice2005} showed $\beta_{crit}\approx 6-7$ for disks with $\gamma=5/3$. A similar value of $\beta_{crit}$ was also
found by \citet{Mayer2005} using a different 3D SPH code
while comparing the evolution of isolated and binary
protoplanetary disks.
However, \citet{Meru2011b} carried out similar SPH simulations to those in \citet{Rice2005} and found non-convergence of the critical cooling rate when
the resolution of the simulations was increased.  They found disks can fragment at a higher $\beta_{crit}$ value in higher resolution simulations. \citet{Meru2012} suggested
that the non-convergent behaviour was caused by extra heating due to artificial viscosity occurring
at low resolution. At higher resolution, less heating is generated through artificial viscosity and a lower  critical cooling rate (i.e. a larger $\beta$) is capable of balancing  the combined heating by gravitational instability and artificial viscosity.  Earlier on \citet{Paardekooper2011} had run 2D simulations with the grid-based code FARGO, showing
that the non-convergence problem is not specific to SPH.
By running a large set of simulations with
both 3D SPH and the 2D grid-based code  FARGO,\citet{Meru2012} concluded that
 $\beta_{crit}$ should converge to an asymptotic value $\approx 20-30$. This has important implications since
 it suggests fragmentation can occur even with relatively
 long cooling times, corresponding to almost an order of
 magnitude longer than the local orbital time. If clumps
 produced by fragmentation  can later evolve into gas giant
 planets (see eg. \citet{Helled2014}), it follows that giant planet formation via disk instability is more common than
 previously thought.

\citet{Paardekooper2011} suggested that the non-convergence is at least partly due to the emergence of special locations in the disk, at the boundary between the turbulent and the laminar region, which
appear when the disk is still adjusting towards a 
well developed gravito-turbulent state.
In order to avoid this edge effect, \citet{Paardekooper2012} carried out high resolution 2D local shearing sheet FARGO simulations, but found a similar increase of $\beta_{crit}$ as resolution increased. Fragmentation also exhibited a stochastic nature in his simulations. However,
recently \citet{Klee2017} performed a numerical study
arguing that stochastic fragmentation is probably caused by oversteepening when performing slope limiting in grid-based
codes.
Moreover, \citet{Young2015} showed that the fragmentation boundary is strongly related to the gravitational softening in 2D simulations. When the softening is comparable to the resolution scale $\beta_{crit}$ increases with resolution, while when the softening is comparable to the disk scale height, $\beta_{crit}$ varies little. \citet{Young2015} also showed that clumps form from overdensities with length-scales $\sim H$. This is consistent with
the earlier result of \citet{Gammie2001}, who found no power at scales much smaller than $H$, which were resolved in the simulations. These results combined suggest that, whatever
is the nature of non-convergence, it should involve
physics and/or numerics acting on a scale of order $H$.

In this paper, we attempt to explore again the issue of 
non-convergence by comparing different Lagrangian hydro
methods which have by design different numerical dissipation. We are driven by the notion that non-convergence might be caused by numerical errors, 
perhaps associated with the boundary regions identified
by  \citet{Paardekooper2011}. Our aim is to identify
the exact source of non-convergence as well as a way
to overcome the problem, in order to place self-gravitating
disk simulations on a firm ground.

We run simulations using exactly the same initial condition of \citet{Meru2012} with three different Lagrangian hydro methods.
Specifically, we use
the vanilla SPH implementation in the GIZMO code \citep{Hopkins2015}, which is equivalent to the original GADGET3 code (\citep{Springel2005} with standard Monaghan
artificial viscosity and Balsara switch \citep{Balsara1995}, 
a modified version of the same code where we implemented
the  artificial viscosity scheme of \citet{Cullen2010}
(inviscid SPH, hereafter ISPH), and the 
new Lagrangian meshless finite mass(MFM) method
in GIZMO \citep{Hopkins2015},which does not need any
artificial viscosity.
Finally, a smaller set of additional simulations exploring 
recent variants of the SPH method are presented at the end
of the paper.

In section 2, we describe the setup of the simulations. In section 3, we present the main results. We discuss the 
interpretation and implications of our results in section
4 ,and finally draw our conclusions in section 5. Results
of ancillary numerical tests are also presented in Appendix
A and Appendix B.

\section{Simulations}

In order to compare with previous work, we use the same disk model as in \citet{Meru2012}. A $0.1 M_{\odot}$ disk spans a radial range, $0.25<R<25$ AU, surrounding a solar mass star. The initial surface mass density and temperature profiles are $\Sigma \propto R^{-1}$ and $T \propto R^{-1/2}$, respectively. The temperature is normalized so that at the outer edge of the disk the Toomre $Q$ parameter equals 2. We use an adiabatic gas equation of state with $\gamma=5/3$. We note that the problem is essentially scale free and we 1 solar mass, $1$ AU, and $1$yr as the mass, length and time unit respectively in our calculation. The simulations are run with different cooling rates as well as 
three different resolutions. The low resolution model (LR) comprises 
250,000 particles,  while 2 million and 16 million particles  are used, respectively, in the high resolution (HR) and ultra-high
resolution model(UHR).
The central star is modeled with a sink particle with sink radius of 0.25 AU. Gas particles reaching the sink 
radius are deleted and their mass  and momentum are added to it to ensure mass and momentum conservation.
We use adaptive gravitational softening\citep{Price2007}.

 The three disk models are run with the three different hydro-methods, namely standard  vanilla SPH as
in  \citet{Meru2012} (hereafter TSPH),  inviscid SPH (ISPH),
which uses the implementation of artificial viscosity
by Cullen \& Dehnen \citep{Cullen2010}, and  the MFM method.
The first two methods differ only in the way artificial
viscosity is implemented. In vanilla SPH we solve the
fluid equations using the standard density-entropy formulation\citep{Springel2005} and standard Monaghan artificial viscosity
with Balsara switch \citep{Balsara1995}, designed to reduce
viscosity in flows with high vorticity. In ISPH we use the same formulation of the hydro equations but we use the Cullen \& Dehnen viscosity 
which uses the time derivative of the velocity divergence 
as shock indicator in order to eliminate numerical viscous
dissipation away from shocks. In this formulation, particles have individual adjustable viscosity coefficients
between an $\alpha_{max}$ and an $\alpha_{min}$ value,and are
further assigned a coefficient $l$ that sets the decay time
of viscosity away from shocks.Therefore the latter formulation requires to set three parameters instead
of the conventional two coefficients $\alpha_{sph}$ and $\beta_{sph}$ in the standard Monaghan viscosity. While
previous artificial viscosity schemes had appeared that
used switches acting on individual
particles \cite{Morris1997}, including with similar parameterization
\citep{Rosswog2000}, the ISPH method is a further improvement towards a truly inviscid
method owing to its shock tracking
procedure.
The MFM method is instead an entirely different method to
solve the hydrodynamical equations, but shares with SPH the
Lagrangian approach and the use of particles as tracers of the fluid. Most importantly, it does not need any artificial viscosity in order to generate stable solutions,
even in high Mach number flows, hence it is by nature inviscid. While we defer the reader to \citet{Hopkins2015} for
a detailed description of the code here we recall that
MFM  uses the SPH-like kernel of a particle distribution 
to construct a volume partition of the domain. The fluid
equations are then solved on the unstructured mesh thus
generated by the partition using a Riemann solver (HLLC 
in this paper, see \citet{Hopkins2015}). As the volume elements
are constructed from the particles the smoothing length $h$
of the assigned kernel function still defines the fundamental 
resolution length as in SPH.
The method  conserves mass, momentum and angular momentum by design as an SPH code, but this is even more strictly true for angular momentum relative to SPH since there is no added artificial viscosity, as pointed out and demonstrated with a series of tests in \citet{Hopkins2015}. Being an integral formulation, in this
sense an hybrid with finite-volume grid-based methods, it should also not suffer from the errors associated with poor estimates of gradients proper of SPH.
Gravity is solved using the same tree code in all three methods, inherited from the progenitor GADGET3 code.

For all codes we employ two different techniques to start the simulations. In the first one we start directly from the initial
disk model without any relaxation phase, while in the second one we first relax the system by running the disk for 4 outer rotation periods(ORPs) with relatively long cooling time, using $\beta=12$. By relaxing the initial condition in the second case we allow the 
development of a gravoturbulent state, with sustained spiral structure but no fragmentation, and then switch to the desired  $\beta$ value. The latter setup,
which we indicate with TIC(turbulent initial condition) resembles that adopted by \citet{Paardekooper2011}. It avoids the transition from a smooth disk to a turbulent disk as in figure \ref{f:resl}. We will show this transition can lead to numerical fragmentation, as
also found by \citet{Paardekooper2011}.
For both initial setups, once the desired cooling timescale is turned on, simulations are run either for at least 6 ORPs, or until the disk fragments. 

In order to facilitate comparisons we adopt standard
values of artificial viscosity parameters in both TSPH and ISPH runs.
For TSPH runs we use the Monaghan viscosity with $\alpha_{sph}=1$ and $\beta_{sph}=2$, which is the  default choice for flows in which high Mach numbers can arise \citep{Hernquist1989},
and has been advocated also by \citet{Mayer2004} and \citet{Meru2012}.
For ISPH we set $\alpha_{max}=2$, $\alpha_{min}=0$ and $l=0.05$\citep{Cullen2010}.

We declare a disk to undergo fragmentation (labeled ``F" in  table \ref{t:simulations}) when clumps form
that  are at least two orders of magnitude denser than their surroundings and can survive at least one outer rotation period without 
been sheared apart. Transient overdensities can form in the runs that we label ``NF" but no robust clumps form in the way just defined.
The results of simulations are summarized in Table \ref{t:simulations},  where we indicate whether or not the disk fragments, and for which value of $\beta$.

\begin{deluxetable}{cccccccc}[h!]
  \tablenum{1}
  \tablecaption{Simulations fragmented are marked with F, otherwise are marked with NF. The boundary between fragmentation and non-fragmentation is marked with green color.\label{t:simulations}}
  \tablewidth{0pt}
  \tablehead{
    \colhead{Simulation} &\multicolumn{7}{c}{$\beta=$}\\ & \colhead{3} & \colhead{3.5} & \colhead{4} & \colhead{5} & \colhead{6} &\colhead{8} & \colhead{Boundary}}
  \startdata
  TSPH-LR     &   &    &    & F & \cellcolor{green}  F  & NF & 6-8      \\               
  ISPH-LR     &   &    &    & \cellcolor{green} F & NF &    & 5-6      \\                
  MFM-LR      &  F & \cellcolor{green} F  & NF & NF&    &    & 3.5-4    \\                
\hline
  TSPH-HR     &   &    &    &   &    & \cellcolor{green} F  &  $>$8    \\                 
  ISPH-HR     &   &    &    &   & \cellcolor{green}  F & NF & 6-8      \\                 
  MFM-HR      &   F &  F  &\cellcolor{green} F &  &  NF  & NF   & 4-6    \\               
\hline
  TSPH-LR-TIC &   & \cellcolor{green} F& NF &    &    &    & 3.5-4 \\   
  TSPH-HR-TIC &   &    & \cellcolor{green} F & NF &    &    & 4-5   \\               
  MFM-LR-TIC  &   \cellcolor{green} F & NF &   &   &    &    & 3-3.5    \\               
  MFM-HR-TIC  &  \cellcolor{green}  F &  NF &  &   &    &    & 3-3.5    \\               
  MFM-UHR-TIC &  \cellcolor{green} F  & NF  &  &   &    &    & 3-3.5    \\
  \enddata
\end{deluxetable}

\section{Results}

\subsection{Overview}

Table 1 summarizes our results on fragmentation.
It highlights three main results borne out of our suite of runs. First, the TSPH runs confirm the dependence of the critical $\beta$ for fragmentation on resolution.
Second, we can infer that non-convergence with resolution is strongly mitigated by methods that reduce numerical dissipation in shear flows. This includes improved  SPH artificial viscosity schemes designed to reduce it in shear
flows, as in the case of ISPH, but is even more evident for a new hydro schemes with no explicit numerical viscosity, such as MFM. Third, {\it exact} convergence is obtained only with MFM but demands the
to start from a relaxed, gravoturbulent initial setup, confirming the claim of \citet{Paardekooper2011}
(see Table \ref{t:simulations}, ``TIC" runs). Without relaxation a marginal increase of the critical $\beta$ with increasing resolution persists.   In section 3.2 we will
provide an explanation of these main findings.

At low resolution the fragmentation boundary in TSPH is found to lie between $\beta=6$ and $\beta=8$, which is slightly higher than the fragmentation boundary $\beta=5.5-5.6$ 
in \citet{Meru2011b}. However,in their default setup
\citet{Meru2011b} use $\alpha_{sph}=0.1$ and $\beta_{sph}=0.2$. If we compare instead with the subset
of their runs that adopted $\alpha_{sph}=1$ and $\beta=2$,
consistent with ours, our results are in substantial agreement.
Note that \citet{Meru2012} show that using low values of the coefficients $\alpha_{sph}$ and $\beta_{sph}$ 
can result, counterintuitively, in high dissipation resulting from random particle motion. The noisier shear flow resulting in SPH simulations
with lowered coefficients in the standard Monaghan viscosity scheme had previously been shown to enhance fragmentation in locally isothermal simulations\citep{Mayer2004}.
We also remind the reader that $\alpha_{sph}=1$ and $\beta=2$  are the preferred values for high Mach number flows in a variety of astrophysical regimes, and have
been advocated for both galactic and protoplanetary disk simulations (\citet{Mayer2004, Kaufmann2007}).

\begin{figure*}[ht!]                       
  \plotone{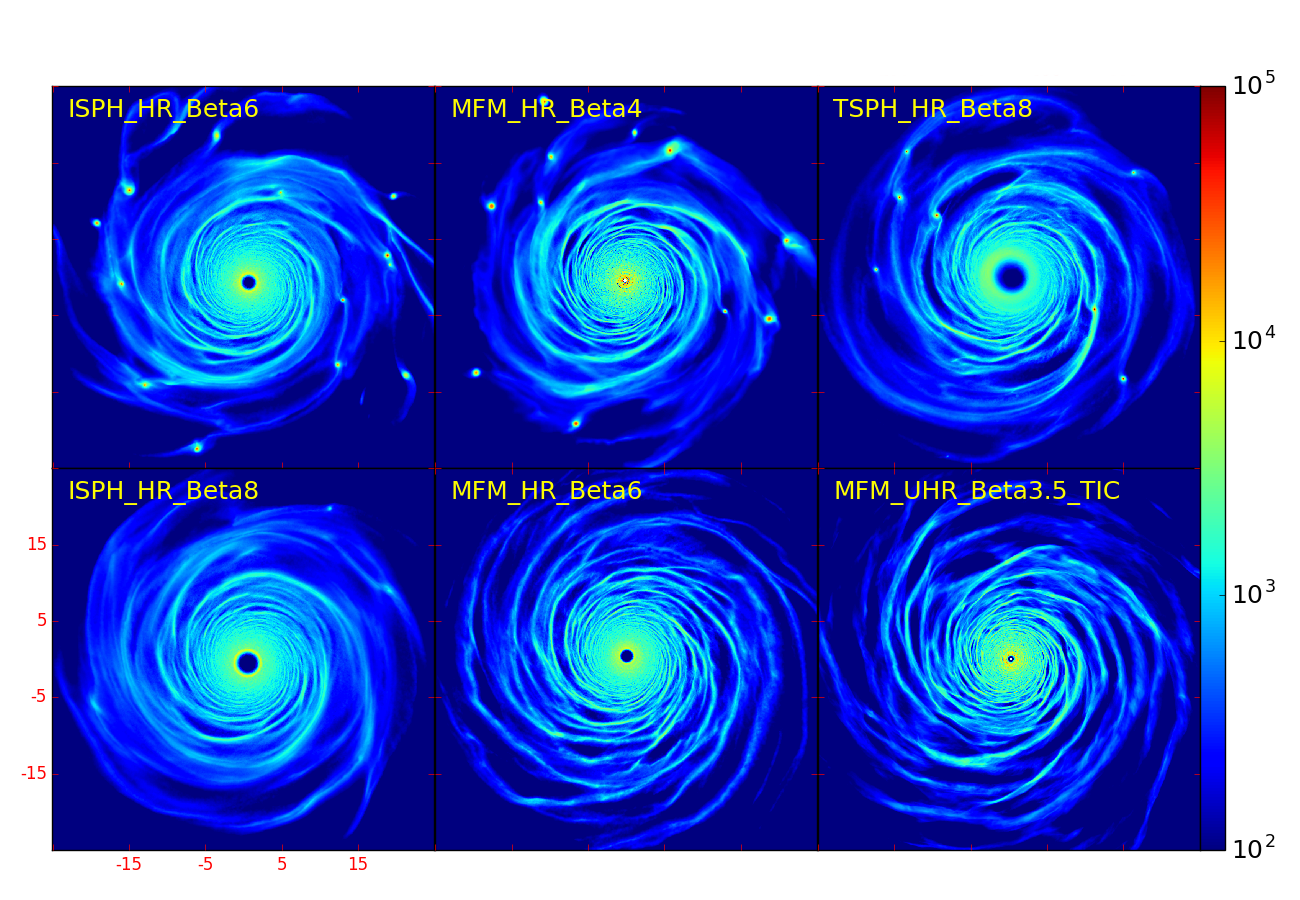}                       
  \caption{Surface mass density ([$g/cm^{2}$]) in logarithmic scale. The box size is 50 code units (axes in code units are shown in red at the bottom left corner).A
  subset of the simulations for the three different methods
  and with $\beta$ near critical is shown to highlight
  that MFM and ISPH do not fragment with much lower values of $\beta$  relative to TSPH at comparable resolution
  (compare bottom panel to top panel).
  A TIC run at even higher resolution  (UHR) is also shown in the bottom right panel, which also does not fragment
  despite a  very low $\beta$ (see Table 1).
  From the first to the second row, left to right, snapshots are taken at 3.7, 0.8, 5.8, 6, 6, 6 ORPs. Note
the different times were chosen for the snapshots in the top panel according to when fragmentation begins, which is much earlier than the time chosen in the bottom panel. The non-fragmenting runs have reached a self-regulated state even before 6ORPs, which is thus a reliable reference time.}\label{f:6panel}                        
\end{figure*}

Figure \ref{t:simulations} shows runs using the three hydro implementations for  $\beta$ values near critical. The markedly different behaviours at
equivalent high resolution is evident. MFM needs $\beta= 4$ or lower to fragment,depending on the initial conditions setup, while TSPH fragments for a $\beta$ twice as large,
and ISPH falls in between. We remind the reader that between TSPH and ISPH the only difference is the implementation of artificial viscosity.

The Figure also shows  how MFM simulations starting from turbulent initial conditions  (TIC) are resilient to fragmentation even with very fast cooling at the highest resolution (UHR, see bottom right panel).
The TIC simulations also exhibit a more flocculent spiral pattern
dominated by lower order modes relative to the non-relaxed
simulations, and this is true for all codes.

\subsection{Numerical dissipation in MFM runs}

When comparing TSPH and ISPH with MFM differences in the hydro implementation are more subtle. Indeed MFM does not have explicit numerical viscosity, rather it solves the hydro equations analogously  to a finite volume method using a Riemann solver on volume elements mapped from the particle distribution. There could be some residual numerical dissipation associated with the non-analytical  Riemann problem solution, but at the same time MFM does not advect fluid elements through a grid, a common
source of numerical dissipation in finite volume grid-based methods. In the Keplerian ring test shown in \citet{Hopkins2015} MFM conserves angular momentum better than all the
other methods it was compared to, including SPH with a modified hydro force (pressure-entropy,hereafter PSPH)
that helps to remove unwanted numerical effects, such as artificial tension, at fluid
interfaces.
This would suggest MFM is inherently less viscous than 
all variants of SPH , irrespective of the artificial viscosity scheme adopted.
However in the latter test the disk is 2D, adiabatic and
non self-gravitating,  which is very different from the configuration  we are studying here. 
Therefore,  in order to reassess that MFM
is indeed less viscous in a relevant albeit simple configuration  we run two different test problems.First
we consider the rotating uniform isothermal sphere collapse test in \citet{Boss1979} . 
The results are shown in appendix \ref{sec:test1}.
We find that angular momentum is transported slightly faster outwards in TSPH runs as expected from stronger numerical viscous transport, albeit differences are small
and overall the angular momentum profiles are comparable
between the two codes at late times.
Furthermore, in appendix \ref{sec:test2}, we use an \emph{accreting} sink particle in  a shearing sheet
configuration. Here the flow responds more abruptly
to the strong gravitational pull of the sink, and 
differences between the three methods emerge more
clearly, establishing that MFM has better angular momentum conservation in flows with large velocity gradients.
While we find that ISPH, as expected, has a lower numerical
viscosity, the accretion rate, as measured by the growth of
the central density, is comparable to TSPH. Instead, MFM
leads to consistently smaller density increase, and the
differences remains over time.

\begin{figure}[ht!]                                                                       
  \plotone{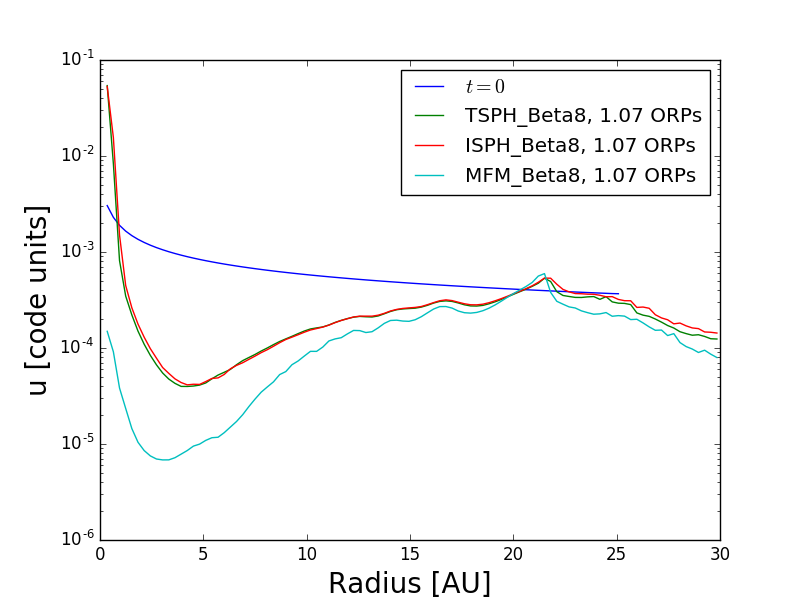}                                                                  
  \caption{Azimuthally averaged radial internal energy 
  profile in the disk mid-plane, in code units and logarithmic scale at 1.07 ORPs. The HR simulations run with different hydro methods are shown. The MFM run has a much cooler disk due to less numerical viscosity and thus reduced numerical heating (see Section 3).}\label{f:u}                                          
\end{figure}

In addition to angular momentum dissipation, the evolution of the internal energy is another important proxy for how strong the numerical dissipation by artificial viscosity is, especially in spiral shocks.
Figure \ref{f:u} shows the internal energy measured at the disk mid-plane after 1.07 ORPs in a non-fragmenting hi-res run (for which we used a large enough $\beta = 8$ to avoid fragmentation).
A non-fragmenting run is a conservative choice  for our purpose since spiral shocks are weaker.
The Figure highlights how the MFM run exhibits
the cooler profile everywhere because, at an equivalent 
cooling rate, it suffers less from artificial viscous
heating, even in the region near the inner boundary 
where SPH is notoriously problematic in handling the
rapidly increasing shear \citep{Mayer2004,Kaufmann2007}. The TSPH and ISPH simulations are almost indistinguishable.
The similar internal energy peak occurs
slightly above 20 AU in all runs because that is where
a strong ring-like overdensity develops (similar to those
in the upper panel of Figure \ref{f:resl}), suggesting
that where the flow is intrinsically more compressive
in all codes the dominant heating comes from PdV work rather than numerical viscosity.

Having now assessed that MFM is the less viscous among the
methods considered, we can turn again to the interpretation
of the results in Figure \ref{f:6panel}.
If, as suggested in \citet{Meru2012}, reduced numerical viscous heating as the resolution is increased is the source of non-convergence, one would expect that MFM runs should be more prone rather than less prone to fragmentation relative to TSPH and ISPH runs. This is because gas can cool more effectively at a given $\beta$,as we have just seen by
comparing the internal energy evolution. Instead the opposite
trend is seen -- MFM is less prone to fragmentation and
its behaviour is nearly convergent with resolution (exactly
convergent in TIC runs). Furthermore, ISPH and TSPH
appear to be equivalent when we inspect the internal energy
evolution in Figure \ref{f:u}, yet ISPH is less prone to fragmentation.
All this suggests that, in order to understand the nature
of our results, among the effects associated with  numerical viscous dissipation, enhanced
angular momentum transport, as opposed to enhanced heating,
is key. We discuss this crucial point, as well as dependence on resolution, in the next section. 

\subsection{Numerical dissipation and the resolution-dependent flow properties}

\label{sec:av}
In standard SPH methods artificial viscosity is needed for the stability of the flow near discontinuities. Without that particle-interpenetration occurs and shocks
cannot be properly resolved. The conventional form of artificial viscosity employs two terms \citep{Monaghan1983,Springel2005};  one term linear in the divergence of the flow, controlled by the $\alpha_{sph}$ coefficient, and one term with quadratic dependence on the flow divergence,  controlled by $\beta_{sph}$. The quadratic term is dominant in shocks. When $\beta_{sph}=0$, the linear term can be recast in 
a shear viscosity $\eta=(1/2)\alpha_{sph} \kappa h c_{s}\rho$ and a bulk viscosity $\zeta=(5/3)\eta$, where $h$ is the SPH smoothing length,
which controls the resolution, and $\kappa$ is a constant
that should be chosen according to the kernel \citep{Cullen2010}. The shear component of the viscosity
will always be present,namely also in non-shocking regions,
and can generate artificial angular momentum transport in
shearing flows such as those inherent to astrophysical
disks, even in presence of damping factors dependent on the
local vorticity such as the Balsara switch \citep{Kaufmann2007}.
\citet{Cullen2010}, building
on previous work \citep{Morris1997, Rosswog2000} 
proposed a new artificial viscosity switch with a high order shock indicator which results, in 
a much more effective damping of shear viscosity away from shocks. 
It is thus expected that, at fixed resolution, ISPH runs, which employ the latter viscosity scheme, should be less affected by numerical dissipation of angular momentum. At the same time, we caution that one
expects mild shocks to  arise as the spiral
pattern grows in amplitude and the disk becomes marginally
unstable, which could then result in significant viscous
dissipation even in ISPH.

Having shown in more than one way, in the previous 
section, that MFM is a less viscous method, and knowing
that ISPH is by construction less viscous than TSPH, we
have inferred that the 
different susceptibility to fragmentation of the three
type of runs must be somehow caused by the different degree
of angular momentum dissipation. 
The next step  is to address how angular momentum
dissipation affects fragmentation. Moreover, there is
a related puzzling fact emerging from Table 1. This is that
one would expect  the results of hydro methods whose
main difference is the viscous dissipation of angular momentum should eventually converge  as the
resolution is increased because,owing to the dependence of shear viscosity on the smoothing length {\it h} just highlighted, viscous dissipation should decrease as resolution increases. Instead the opposite
is seen;$\beta_{crit}$ keeps increasing for TSPH as the
resolution is increased, hence departing more and more
from the $\beta_{crit}$ value determined with MFM. 
However, we caution already at this point that this
conventional way of reasoning is correct only if no 
other property of the flow changes with increasing
resolution. If, for some reason, velocity gradients 
become intrinsically stronger as resolution is increased, as one could imagine that convergence might be hard  to achieve irrespective of resolution.

Here we argue that the \emph{nature of the flow in gravitationally unstable disks does change as resolution 
is increased}. It  develops  regions of stronger shear and
vorticity as sharper flow features become resolved, and such regions then become  sites of higher numerical viscous dissipation. The resolution-dependent nature of the flow also explains why the way the initial setup is prepared matters for the fragmentation outcome {\it at varying resolution}, as we illustrate below.
\begin{figure*}[ht!]                                              
  \plotone{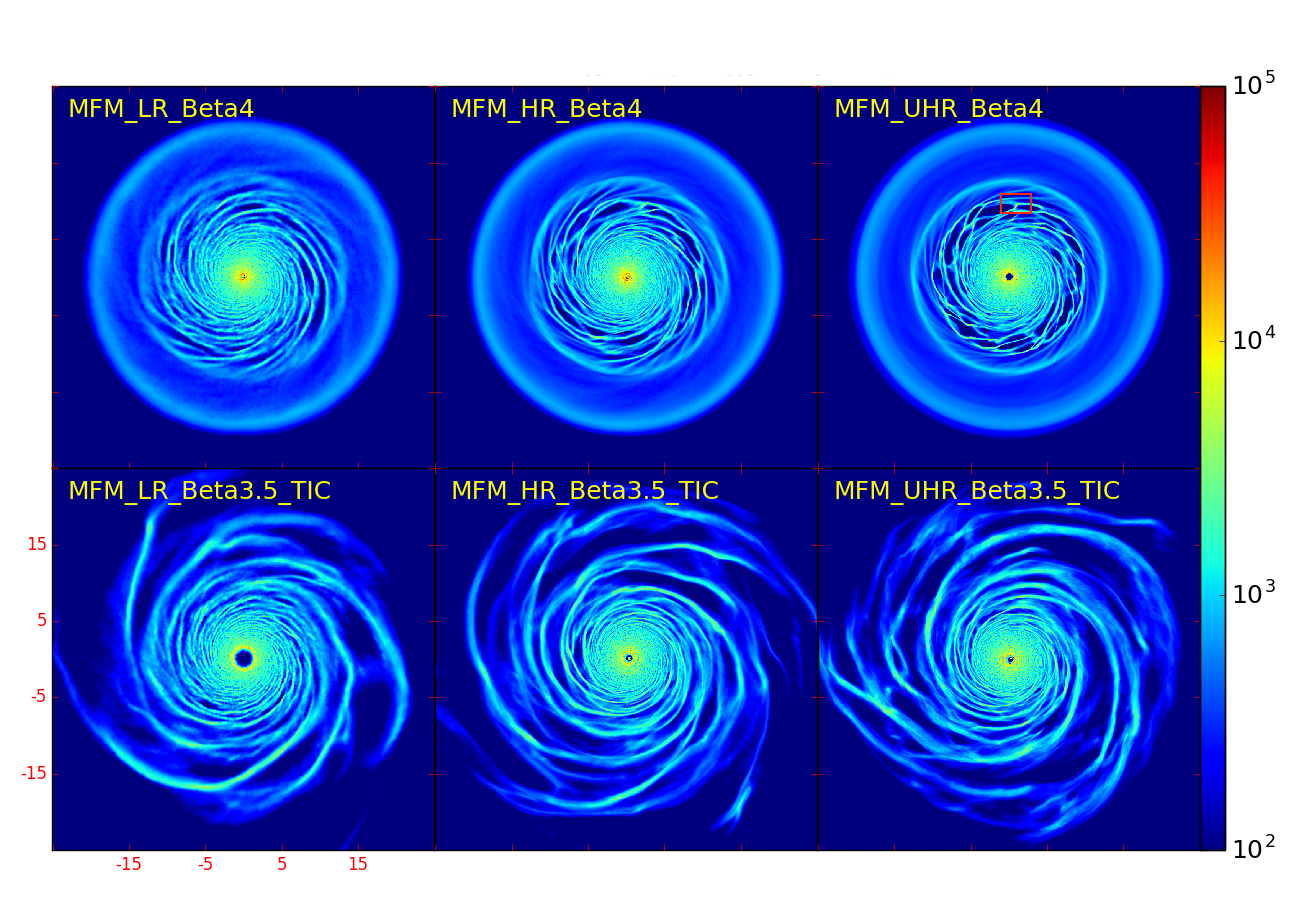}                  
  \caption{Surface mass density ([$g/cm^{2}$]) in logarithmic scale for a subset of the MFM simulations run with different resolution and $\beta$ close to critical
  (axes in code length units are shown in red at the bottom left corner). The bottom panel shows the TIC runs.
  Resolution increases by nearly 3 orders of magnitude from left to right. All snapshots are taken at 1.07 ORPs. A cusp is highlighted in the red box in the top right panel. Cusps form in the interface regions between spirals and outer smooth flow. They cause strong numerical dissipation and are absent in TIC runs (compare bottom and top panels).}\label{f:resl}                    
\end{figure*} 
The changing nature of the flow with resolution is well illustrated in figure \ref{f:resl}. The simulations
using unrelaxed initial conditions offer the best tool
to understand what happens.In these the spiral pattern originates in the inner disk and propagates
outwards\citep{Meru2011a}. We refer to the stage 
before spirals reach the outer edge of the disk as the {\it transition phase}. The duration of such transition phase depends on the cooling rate, but in general lasts  about 2 ORPs.
By visual inspection it is clear that at higher resolution non-axisymmetric modes are better resolved. This is the result of an improved gravitational force resolution, a point already emphasized in \citet{Mayer2004},\citet{Mayer2008} for both SPH and adaptive mesh refinement (AMR) codes. 

\begin{figure}[ht!]                        
  \plotone{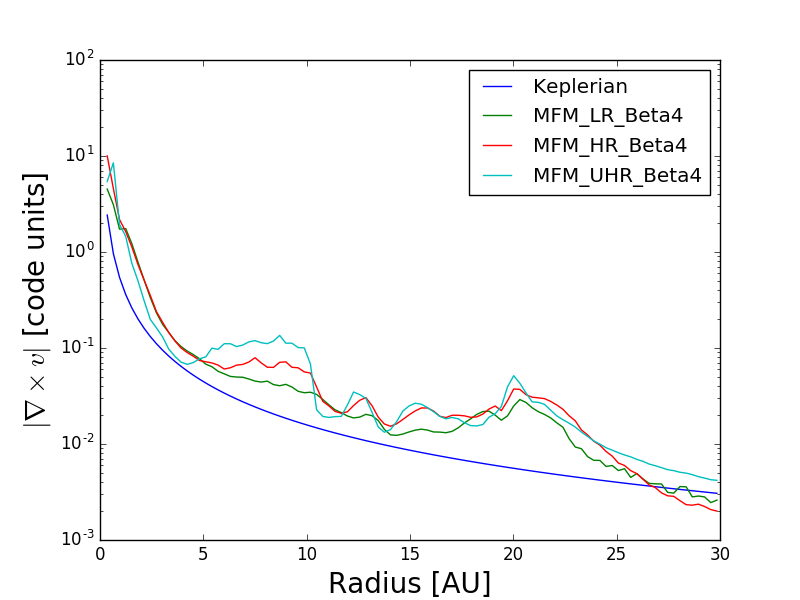}                       
  \caption{Azimuthally averaged radial vorticity profile in the disk mid-plane (the modulus of vorticity is in code units)
  measured at 1.07 ORPs for MFM simulations with different
  resolution.
 The vorticity of a pure Keplerian flow is also shown for comparison}\label{f:vtp}                                   
\end{figure}

We refer to the boundary zones between the  turbulent, highly non-axisymmetric flow component generated by gravitational instability and the background smooth,axisymmetric disk as the {\it interface region}.
As seen in the upper panels of figure \ref{f:resl}, these are regions roughly a few code  units in size, and occur during the transition phase (see the rectangular region in Figure \ref{f:resl}). Afterwards the entire disk
becomes non-axisymmetric and gravito-turbulent.
These interface regions are
relevant because they are the location of the largest local
velocity gradients, associated with both strong shear and vorticity. 

\begin{figure*}[ht!]                                                                      
  \plotone{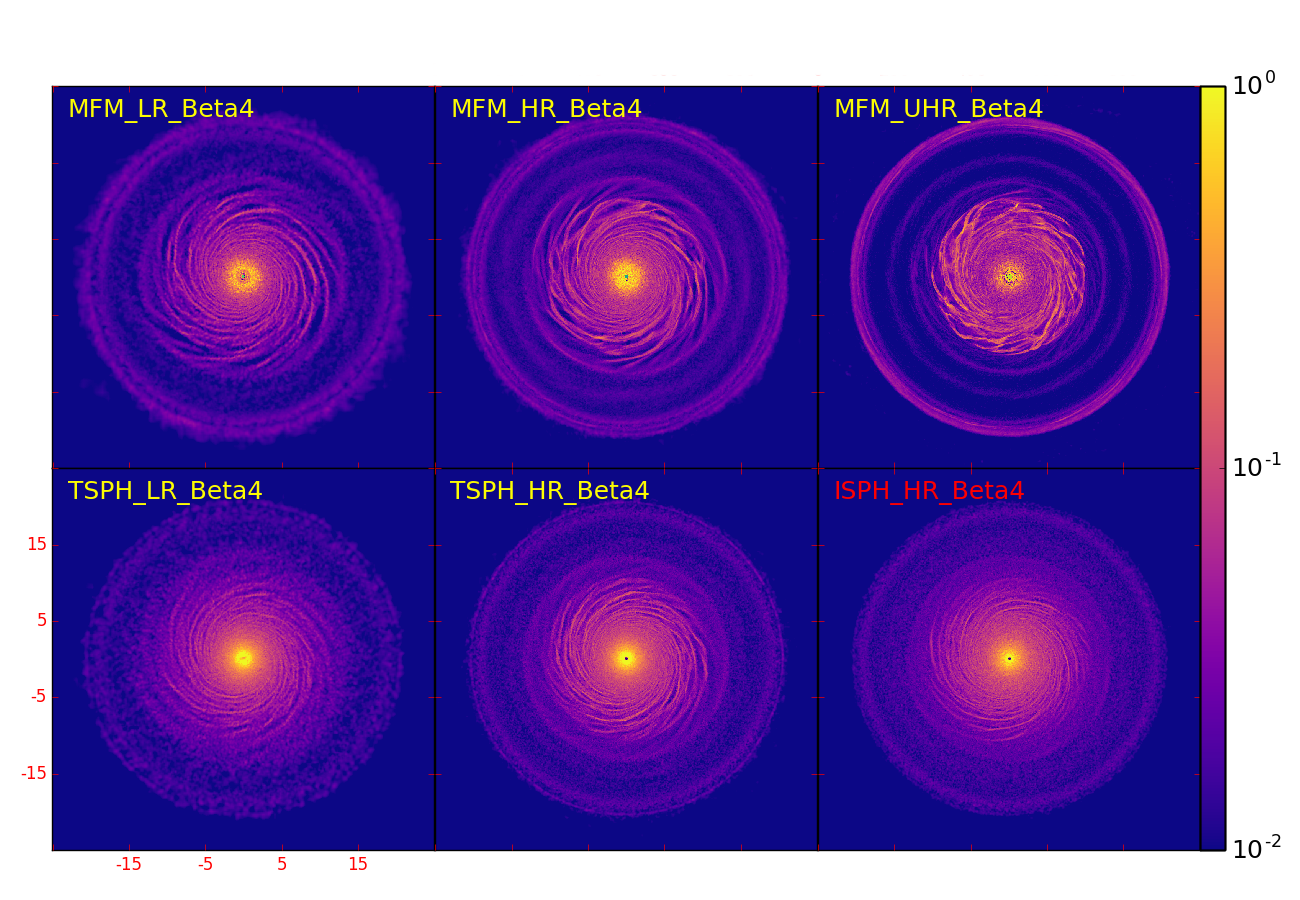}                                                                     
  \caption{Vorticty maps.
We show the modulus of vorticity in the disk mid-plane, in code units (axes in code length units are shown in red at the bottom left corner).
In the innermost regions, vorticity is dominated by the Keplerian rotation, see the vorticity  profile in figure \ref{f:vtp}. In the MFM simulations(top row), as the resolution increases the vorticity due to local
 overdense regions increases, as such regions become
 better resolved and are characterized by strong departures from the  background flow velocity.
 The corresponding peak volume density in code units, in the interface region (located at 8-11 code length units), from left to right, upper to lower, is 0.0026, 0.0081, 0.019, 0.0013, 0.0048, 0.0031.}\label{f:vortm}                                                                
\end{figure*}

We will use the analysis of vorticity as a
proxy for high velocity derivatives in the flow
as vorticity also carries information on the local angular
momentum, which we posit plays a central role.
The azimuthally averaged  vorticity profile in the mid-plane is plotted in figure \ref{f:vtp} for MFM runs at different resolution.
In all of them it is evident that vorticity can be an
order of magnitude higher than the Keplerian flow vorticity. These vorticity peaks occur near interface regions and more pronounced and more frequent as resolution
increases. The latter is an effect of increased force
resolution as higher overdensities appear that
induce higher local vorticity. 
Indeed, if numerical dissipation
of angular momentum does not affect the properties of
the flow one would expect density and vorticity to remain
correlated. Let us consider an approximately spherical {\it cusp} of the flow located at the interface, namely an overdense region of radius $R_c$ and enclosed $M_c$, so that its mean density is $\rho = M_c/{R_c}^3$ . When the cusp density is large enough a fluid element at its boundary  will move with a (circular) velocity $V_c$ such that $\sqrt{G\rho} \propto \frac{V_c}{R_c}$. In other words, the cusp dominates the
local gravitational field, even if collapse has not
ensued, so that locally the vorticity of the flow should
be correlated with local density, since, dimensionally,
the vorticity  $\bm{\nabla}\times \bm{v} \propto \frac{V_c}{R_c}
\sim \sqrt{G\rho}$
($\bm{v}$ is the velocity vector of the gas flow)
We can now turn back to our simulations and check whether
or not vorticity and density in the cuspy regions grow
at the relative rates implied by the proportionality just highlighted. 
First of all, we notice that the
 peak density at interface regions in MFM runs in Figure \ref{f:resl} is almost an order of magnitude larger in the UHR simulation relative to the LR simulation. As expected, vorticity is
 correspondingly higher (Figure \ref{f:vortm}).
 Note that TSPH runs share the same trend. ISPH runs yield a similar vorticity map as TSPH
 because at interface  regions the flow has high velocity derivatives
 hence artificial viscosity is not suppressed by the Cullen \& Dehnen switch.

\begin{figure*}[ht!]                        
  \plottwo{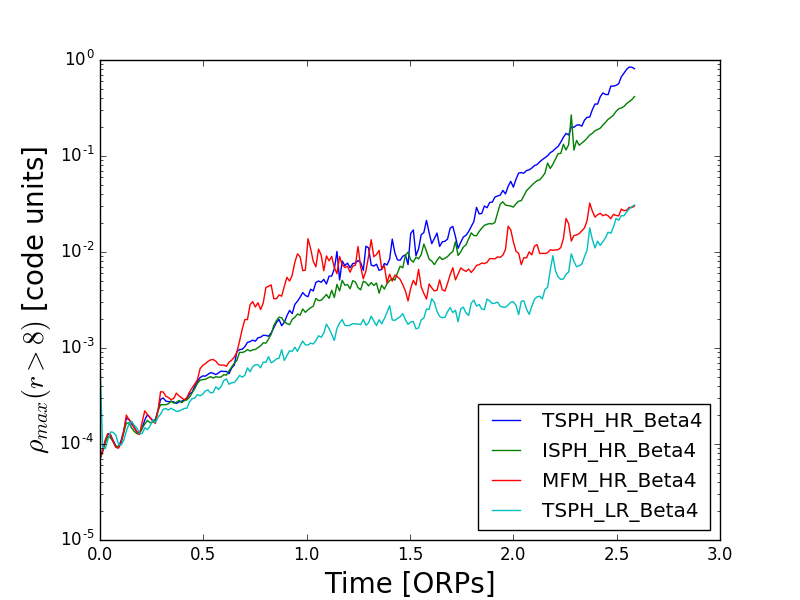}{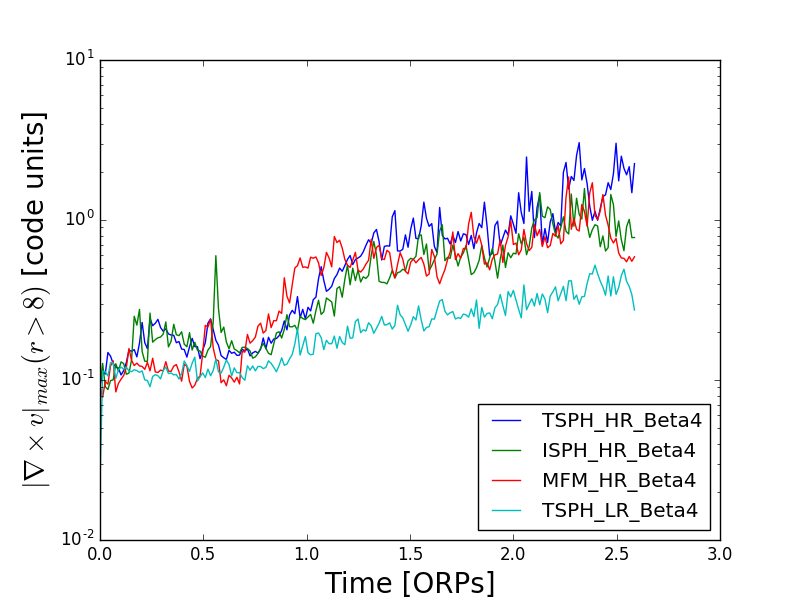}                        
  \caption{Peak volume density(left) and peak vorticity(right) at $r>8$, which includes  the
  interface region, in the first 2.7 ORPs.  We show
  results for TSPH, ISPH ad MFM runs at different resolutions.
  The peak density of SPH simulations grows exponentially after the spirals reach the outer edge of the disk at about 2 ORPs. The peak vorticity is uncorrelated with the peak density, see discussion in section 3}\label{f:pd}
\end{figure*}

We cannot compare the vorticity in the MFM and SPH simulations in figure \ref{f:vortm} directly because the disk is much cooler in MFM(see figure \ref{f:u}) so that the spiral pattern evolves faster than in the SPH simulations. 
But we can still compare 
the evolution of maximum density and maximum vorticity
in a relative sense. We focus on the transition
phase of a set of fast cooling simulations with $\beta=4$.
We plot the peak volume density and peak vorticity in the $r>8$ region in figure \ref{f:pd}. At $t=2 $ ORPs the spiral pattern reaches the outer edge of the disk in all simulations. In SPH simulations very compact, dense clumps have formed at this point while in MFM only loosely bound overdensities are present.
The peak density in MFM is indeed lower than in SPH
simulations at comparable  resolution. Also, SPH simulations show an exponential growth of the peak density which is absent in the MFM simulation.
The figure also shows that, in the early stage of the transition phase, the MFM run experiences a faster growth of density perturbations, reflecting the lower temperature (see Figure \ref{f:u}).
Yet the key result emerges when we compare density and 
vorticity evolution in the different runs. It is noteworthy
that in MFM  runs vorticity appears to grow at the rate expected
based on the relation $\bm{\nabla}\times \bm{v} \propto
\sqrt{G\rho}$.
Indeed at 2 code units it has grown by about an order of magnitude while
density as grown by nearly two orders if magnitude.Conversely,
a comparable increase in vorticity occurs in TSPH and ISPH runs at this time but there density increases much more, by about four orders of magnitude.
We argue that this remarkable mismatch reflects strong numerical dissipation of angular momentum near cusps,which damps vorticity and promotes collapse.

.

\begin{figure*}[ht!]                                                                      
  \plotone{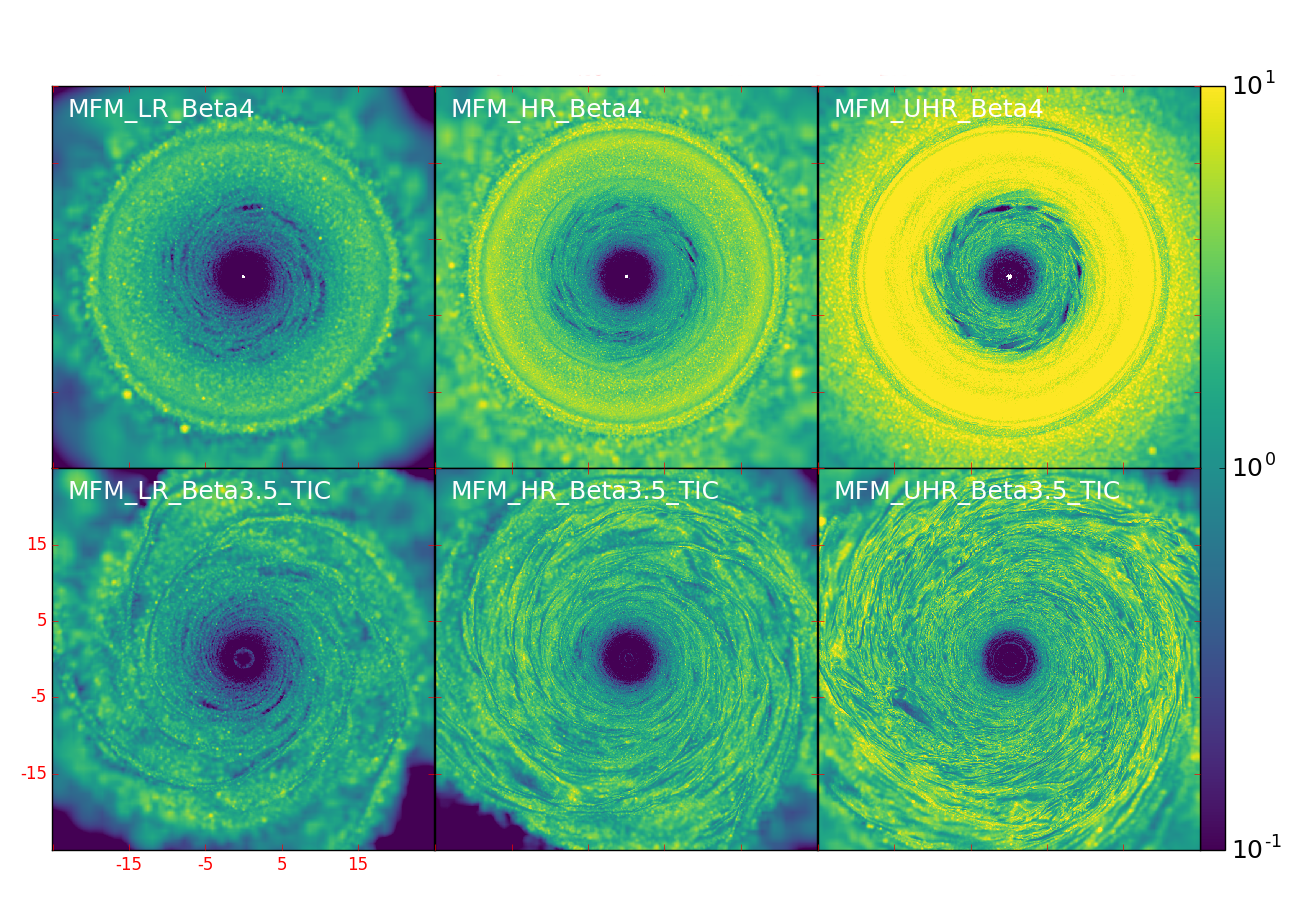}                                                                    
  \caption{The $q$ factor at 1.07 ORPs in MFM runs at different resolution. The top panel shows the standard runs, while the bottom panel shows the TIC runs.
  As the resolution increases $q$ increases mainly because of a smaller smooth length. 
  However, in the interface regions, especially between spirals, $q$ can become very small which indicates strong numerical dissipation (see dark spots, even more prominent at higher resolution in the top panel).
  These regions are not present in the TIC runs, where correspondingly $q$ is higher on average and increases
  with increasing resolution everywhere, in sharp contrast
  with the runs in the top panel.
  The innermost regions have small $q$ due to strong background shear but are Toomre stable
  and axisymmetric so our arguments in Section 3 do not
  apply.}\label{f:qparam}                                                         
\end{figure*}

We can attempt to understand the nature of the problem
by considering resolution requirements in self-gravitating
disks.
In order  to properly follow the fragmentation, the most unstable Toomre wavelength $\lambda_{T}$ must be  resolved\citep{Nelson2006}. This is
\begin{equation}\label{e:lambda}
  \lambda_{T} = \frac{2c_{s}^{2}}{ G \Sigma}
\end{equation}
where $\Sigma \sim \rho H$ and in a marginally unstable disk, ($\Sigma$ and $H$ are, respectively, surface density
and disk scale height). Note that the most unstable
wave number $\frac{2 \pi}{\lambda_T}$ is exactly the
inverse of the disk thickness, so that, dropping constant
factors we can write
$ \lambda_{T} \sim Q {\frac{c_s}{\Omega}} \sim H$\citep{Cossins2009}. Since $Q \sim 1$ when
the disk becomes unstable, for a Keplerian disk, we substitute $\Sigma$ with $\rho \lambda_{T}$ in equation \ref{e:lambda}. We can then write $\lambda_T \sim  \frac{c_{s}}{\sqrt{G\rho}}$.
We recall that the Toomre wavelength is borne out of
linear perturbation theory for axisymmetric local
perturbations. It is a conservative resolution marker 
for our purpose since it has been shown that the nonlinear
stage of fragmentation in spiral arms occurs on a 
characteristic wavelength that is 5-6 times
smaller than the Toomre wavelength \citep{Boley2009,Tamburello2015}. In the interface
regions the spiral pattern dominates the flow hence the
same considerations would apply.
Recalling again that, near cusps, we can write $\sqrt{G\rho} \propto \frac{V_c}{R_c}$ and 
 $\frac{V_c}{R_c}\sim \bm{\nabla}\times \bm{v}$, we
obtain:
\begin{equation}
  \frac{\lambda_{T}}{h}\sim \frac{c_{s}}{h|\bm{\nabla}\times \bm{v}|}
\end{equation}
We define
\begin{equation}
 q \equiv \frac{c_{s}}{h|\bm{\nabla}\times \bm{v}|}
\end{equation}
where $h$ is the kernel smoothing length.
Since the Toomre wavelength should be resolved by at least one kernel\citep{Nelson2006}, the resolution will not be adequate
if $q< 1$ bearing in mind that the Toomre wavelength is
a generous estimate for the characteristic wavelength of 
fragmentation. Likewise, the equation above highlights how, as vorticity increases, the resolution has to increase proportionally, namely the smoothing length $h$ has to decrease, in order to maintain $q$ large and thus
follow a self-gravitating fluid appropriately.
%In passing we  note  that \citet{Cullen2010} also points %out numerical dissipation can be significant if $c_{s}/h$ %is smaller than shear.
The upper panel of Figure \ref{f:qparam} shows that, as we expected,
$q < 1$ in the interface regions, and decreases as resolution is increased, reflecting the dominant effect
of increasing vorticity relative to the smoothing length
decrease.

Conversely,outside these regions
$q$ increases as resolution is increased,as the dominant
effect is now the decrease of the smoothing length. 
Note that Figure \ref{f:qparam} uses the MFM runs.
The lower panel of Figure \ref{f:qparam} also shows that
$q$ is always high {\it at all resolutions} when
the initial conditions are relaxed (TIC runs),which 
reflects the fact that, even for MFM, only in this case
there is exact convergence for the critical
cooling time (see Table 1).

Indeed in TIC runs the velocity field of the flow varies more smoothly across the disk, with no transition phase
and thus no cusps (Figure \ref{f:resl}). This confirms the claim of \citep{Paardekooper2011} on the
importance of the initial setup for convergence studies.
This is because artificial angular momentum transport during the transition from a smooth disk to a fully developed spiral pattern 
is avoided.
Yet table \ref{t:simulations} also shows that only MFM
shows exact convergence among TIC runs ($\beta_{crit}=3$).
The SPH runs still have excessive viscous dissipation of
angular momentum even in presence of more moderate
velocity gradients in shearing flows. While we did not
run ISPH with TIC conditions, we have already shown that the viscosity switch alone becomes rather ineffective at
high resolution, resulting in a behaviour similar to 
TSPH (Figure \ref{f:pd}).

\section{Discussion}

Fragmentation in self-gravitating disks is known to be a process highly sensitive to the numerical implementation of the hydrodynamical  equations  and resolution. Earlier work carried out with both grid-based and SPH codes highlighted the importance of numerical approaches,including the
effect of artificial viscosity, for the simple case of locally isothermal disks \citep{Pickett2001, Mayer2004,Durisen2007,Mayer2008}.
Strong dependence on numerics has also been seen in cloud fragmentation simulations. For example, \citet{Bate2011} carried out SPH simulations of molecular cloud collapse to form protostars and disks, and found more fragmentation with increasing resolution. \citet{Hu2014a} ran galaxy simulations with the SPH code GADGET \citep{Springel2005} and showed different artificial viscosity switches can lead to very different fragmentation results.

In this paper we have shown that simply increasing the
resolution does not yield a more correct answer for
disk fragmentation, rather the nature of viscous dissipation and how it couples with the properties of the
flow is the crucial aspect. Since the properties of the
flow change with resolution, studying convergence as a
function of resolution alone is insufficient, Instead
comparing different hydro methods, and in particular 
different methods with varying numerical dissipation, reveals the real nature of the problem. Angular momentum dissipation in strong shearing regions, which we study using vorticity, is the main reason behind non-convergence in SPH methods.
ISPH,which suppresses artificial viscosity in pure shearing flows, improves considerably but does not fully converge,
likely because in regions of high vorticity the flow
is also highly compressive so that viscosity turns on
based on the shock tracking scheme.

MFM is the only method for which we were able to prove convergence of the critical cooling timescale among the methods  explored so far, but even in this case
exact convergence requires relaxation of the initial
conditions  in order to avoid transients that lead to sharp flow velocity gradients and high local vorticity at 
interface regions.
The important role of the initial conditions is supported also by a recent study of \citet{Backus2016} with locally isothermal disks using a modern formulation of the SPH hydro force, but still standard artificial viscosity, using the ChaNGa code. It is also in line with previous findings by \citet{Paardekooper2011} with the FARGO code.

Clearly we have only considered a small set of hydro methods in this paper.Even in the context of SPH many
different revised formulations of the hydro force have
appeared in the literature over last few years which would
be worth exploring, such as SPHS \citep{Hayfield2011},
GDSPH \citep{Keller2014,Tamburello2015} and 
the, pressure-entropy formulation(PSPH) in \citet{Hopkins2013a}. Thermal diffusion alone has been
shown to be effective at removing artificial surface
tension \citep{Price2008,Shen2010}.It may play a role as it does in all problems
where sharp fluid interfaces appears \citep{Agertz2007}.
These modifications of other aspects of the SPH formulation
could also affect $\beta_{crit}$. The smoothed cooling
approach of \citet{Rice2014} is also another example of
how the fragmentation problem is sensitive to the details
of the numerical implementation within the same category
of methods.
In order to at least partially address this we 
 rerun the HR simulation from relaxed initial conditions
 (TIC) using PSPH and thermal diffusion in the GIZMO code
 \citep{Hopkins2015}, and with the Cullen \& Dehnen viscosity
 formulation. We find that the disk
 fragments for $\beta=4-5$. Therefore using PSPH does not
 lead to an improvement relative to SPH, confirming the
 central role of residual viscous dissipation as opposed
 to other aspects of the SPH scheme, for example the SPH ``E0" zeroth-order errors\citep{Readd2010}.

\citet{Dehnen2012} showed that the kernel functions does affect numerical dissipation in shear flow simulations.The Wendland C4 kernel shows the best conservation properties in their numerical tests. Although MFM solves the fluid
equations on the volume elements constructed from the
particle distribution, the volume elements themselves are
constructed using a specified kernel function hence it is
relevant to ask what effect the choice of the kernel function has on the flow solution.
Therefore we rerun the MFM-HR-Beta3-TIC simulation with Wendland C4, using 200 neighbours.  We found that the simulation still fragments at $\beta_{crit} =3$. This
is reassuring as it strengthens the conclusion that 
the results have truly converged,namely 
we cannot push $\beta_{crit}$ even further by reducing
numerical dissipation using a more conservative kernel function. However, a word of caution must be said.
With such many neighbours the effective gravitational force resolution is half the force resolution in the standard HR simulation, being equal to that in the LR simulations,  and the force resolution at later stage is also not directly comparable  due to its full adaptive nature. 

%We didn't run PSHP simulations with Wendland C4 kernal %because it is hard to tell the decrease of effective force %resolution or better conservative property of the kernel %changes $\beta_{crit}$, if it does change with this %kernel. The $\beta_{crit}$ in MFM simulations truely %converge regardless of the kernel.

\citet{Meru2012} could not prove exact convergence for a set
of FARGO simulations, for which they also considered 
varying the coefficient of viscosity. Indeed, despite
showing a trend with resolution that was pointing
towards convergence, they did not
find $\beta_{crit}$ to maintain the exact same value when they increased their resolution by a factor
of 2. The asymptotic value of  $\beta_{crit}$  suggested
by their results was also  higher than the corresponding value found in their SPH runs. In
light of our findings that show how convergence occurs
to a very small vaue of $\beta_{crit} =3$,we argue that the published FARGO simulations might suffer from even stronger numerical dissipation than SPH, resulting not only
from numerical viscosity but also from advection errors
occurring in cuspy regions that are not at all axisymmetric
and hence cannot be captured properly by the cylindrical geometry of the grid. 
For the same reason, significant improvements can be expected using relaxed initial conditions as deviations
from axisymmetry are less severe due to the absence of
the transition phase. This is once again in line with the
findings of \citep{Paardekooper2011}
Advection errors  in the interface regions could be
even more severe in Cartesian grids,unless aggressive
AMR is used to better capture the curvature of the flow.
We plan to reassess this in a future comparison employing
more than one type of grid-based code.

Finally,at fixed smoothing length the density increase
in self-gravitating flow is strongly affected by gravitational softening. Adaptive softening is expected
to play a key role as it increases the force resolution as particles converge towards a cusp or sharp interface.
We run additional TSPH simulations with fixed softening to
quantify the importance of the choice of softening for the
problem under study.

\begin{figure}[ht!]
  \plotone{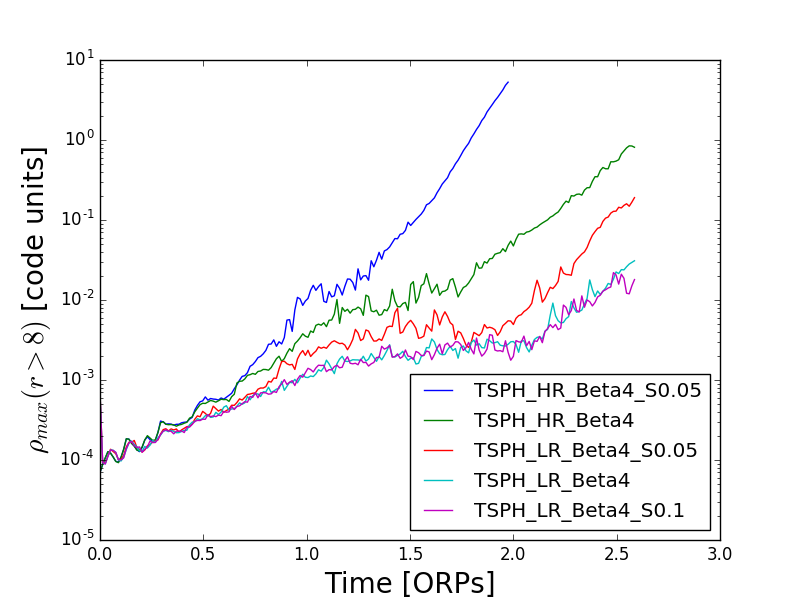}
  \caption{The effect of fixed gravitational softening and resolution on peak density inside the interface region
  at  $r>8$ during the transition phase. We show the 
  results for a set of TSPH runs, including the reference
  runs with adaptive softening.
  The HR simulation with softening 0.02 is stopped early because of small time steps.}\label{f:soft}
\end{figure}

Figure \ref{f:soft} indicates that  also fixed softening simulations also experience an exponential peak density growth.   In the TSPH-LR run, when we choose a gravitational softening of 0.1  
code units, which equals half the smoothing length in the disk middle plane at $r=9$ (\
see the curve of the TSPH-LR-Beta4-S0.1 run), the peak volume density evolution is similar
to that in the the adaptive softening run. If we decrease the softening to 0.05(TSPH-LR-Beta-S0.05)
, the peak volume density increase significantly. The TSPH-HR-Beta4-S0.05 has even higher
 peak density than TSPH-LR-Beta4-S0.05, and also
 even higher than the peak density in the adaptive softening run, resulting from the denser cusps forming
at higher resolution. As a result, 
runaway fragmentation in presence of vorticity damping
can be even
more problematic than in the reference adaptive softening run.

The results found in this paper suggest that disks need to cool really fast to fragment, on a timescale nearly identical to the local orbital time. This reinforces
the notion that it is in the outer disk regions, at
$R > 30$ AU \citep{Rafikov2009,Clarke2009}, that the conditions are
favourable for fragmentation.
However, our study focused on idealized isolated disk
models,neglecting the effect of both external triggers, such as mass accretion onto the disk from the envelope in the early stages of disk evolution
\citep{Boley2010,Vorobyov2010} or internal triggers such as perturbations by a previously-formed fragment \citep{Meru2015}. Mass loading has already been shown to bring disks to fragmentation even when radiative cooling is slow \citep{Boley2009}, or even formally absent  as in the collapse of
polytropic turbulent cloud cores \citep{Hayfield2011}. . The correct treatment of radiative cooling and heating originating from accretion ultimately requires  radiative transfer to be employed (eg \citep{Meru2011a,Rogers2011,Mayer2016} However, attention should be paid to the initial condition setup and accretion boundary condition. Complex flow structures and overdensities arising in turbulent collapse may lead to strong numerical dissipation and cause spurious fragmentation for the same reasons described in this
paper. Therefore a weakly dissipative method such as MFM holds promise to be most suitable for radiative disk formation studies as well.

The complexity inherent to model disk fragmentation emerging from our study suggests that it is too premature
to use existing simulations to compare with exoplanet
detection via imaging surveys for wide orbit gas giants
(eg \citep{Vigan2017}). Indeed, once the formation stage
of clumps is robustly modeled, their evolution via 
further collapse \citep{Galvagni2012, Szulagyi2017} and migration in the disk \citep{Baruteau2011,Malik2015} is also not completely understood, and not yet firm from the numerical side.
Fast inward migration of clumps is routinely found which
alone would invalidate any direct comparison with wide
orbit gas giants as a way to infer the role of disk instability, but the migration rate in self-gravitating disks  might be affected directly or indirectly by numerical dissipation effects.
A comparison of hydro methods is warranted in this case
as well.

Finally, there is an interesting upshot of our study. This
is that
actual physical sources of viscosity in disks, such as the
magnetorotational instability (MRI) \citep{Balbus1991}might promote fragmentation if they can dissipate angular momentum 
efficiently in overdense regions with strong shear.
This will be explored in future work.

\section{Conclusions}
We performed 3D hydrodynamic simulations using SPH with different artificial viscosity implementations as well as with a new Lagrangian method, MFM, which employs a Riemann solver on a volume partition generated by particles, thus not requiring any explicit numerical viscosity.

Our TSPH simulations agree well with a previous study by \citet{Meru2011b, Meru2012}.They confirm the \emph{non-convergence} of the critical cooling rate for disk fragmentation in a self-gravitating disk. MFM instead
attains convergence, and to a significantly lower value
of the critical cooling time, $\beta_{crit} = 3$. ISPH 
simulations, which adopt the more conservative artificial
viscosity implementation by Cullen \& Dehnen \citep{Cullen2010}, resulting
formally in no shear viscosity, exhibit an intermediate
behaviour. Finally,exact convergence, even in MFN, occurs only with relaxed initial conditions.
We find a coherent explanation for our results,showing that
the driving effect is numerical dissipation of angular 
momentum in strong shearing flows. This is exacerbated
as resolution is increased, despite shear viscosity in
SPH is formally decreasing with resolution,because the
flow develops stronger cusps with large velocity derivatives where vorticity is artificially damped. In those regions the characteristic wavelength associated with
fragmentation, conservatively measured with the Toomre
wavelength, remains poorly resolved even when the number of
particles is  increased by nearly three orders of magnitude, hence the high sensitivity on the numerical
dissipation.
Relaxed initial conditions exhibit a more convergent behaviour because they avoid a transient phase in which
overdense cusps and associated high velocity derivatives occur, thus suffering less the impact of artificial angular
momentum dissipation by numerical viscosity.

We stress that the small value  $\beta_{crit} = 3$ in the converged MFM simulations
is significantly smaller than any value previously found
in the literature when trying to assess convergence with
increasing resolution. This indicates that the trends suggestive of asymptotic convergence reported in the literature for
both SPH codes and FARGO \citep{Meru2012,Rice2014} were just reflecting saturation of numerical angular
momentum dissipation rather than a decreasing numerical dissipation with increasing resolution.

We have further assessed the validity of our statements
by running additional simulations that explore other
aspects of SPH implementations,such as adaptive softening,
kernel function and form of the hydro force. These tests
confirm  our claim that only MFM attains exact convergence.

In summary, our results indicate that the use of hydro methods with minimal numerical viscosity is the first necessary step towards predictive simulations of disk instability because the issues we highlighted would still play a role in complex simulations with radiative transfer and envelope accretion.
\newline 

We thank Jim Stone, Frederic Masset for useful discussions. We are grateful to Philip Hopkins for helping run the code. Pynbody\citep{pynbody} is used for the visualization. We acknowledge support from the Swiss National Science Foundation via the NCCR PlanetS. F.M. acknowledges support from The Leverhulme Trust and the Isaac Newton Trust.

\bibliographystyle{aasjournal}
\bibliography{references}

\appendix

\section{Isothermal sphere collapse}
\label{sec:test1}
To further assess the low numerical viscosity in MFM we run the standard isothermal test case for the collapse of a rotating molecular cloud. This test, first described by \citet{Boss1979}, is a typical test case for numerical codes studying fragmentation\citep{Bate1995}. An initially isothermal, spherical, self-gravitating hydrogen molecular cloud starts to collapse from rest. Physical viscosity and the magnetic field are not included, so the angular momentum of the system should be conserved. The gas is assumed to be ideal gas and initial temperature is $10K$. The mass of the cloud is $1M_{sun}$, and radius of the cloud is $3.2\times 10^{16}cm$. The angular velocity is $\Omega=1.6\times 10^{-12} rad s^{-1}$. The ratios of thermal energy to the magnitude of gravitational energy, and rotational energy to the magnitude of gravitational energy of $\alpha =0.25$ and $\beta=0.20$, respectively. The mean density is $\rho=1.44\times 10^{-17} g cm^{-3}$. No density perturbation is imposed initially. The free-fall time for the initial cloud is $t_{ff}=5.52\times 10^{11} s$.

\begin{figure*}[ht!]                                                                      
  \plotone{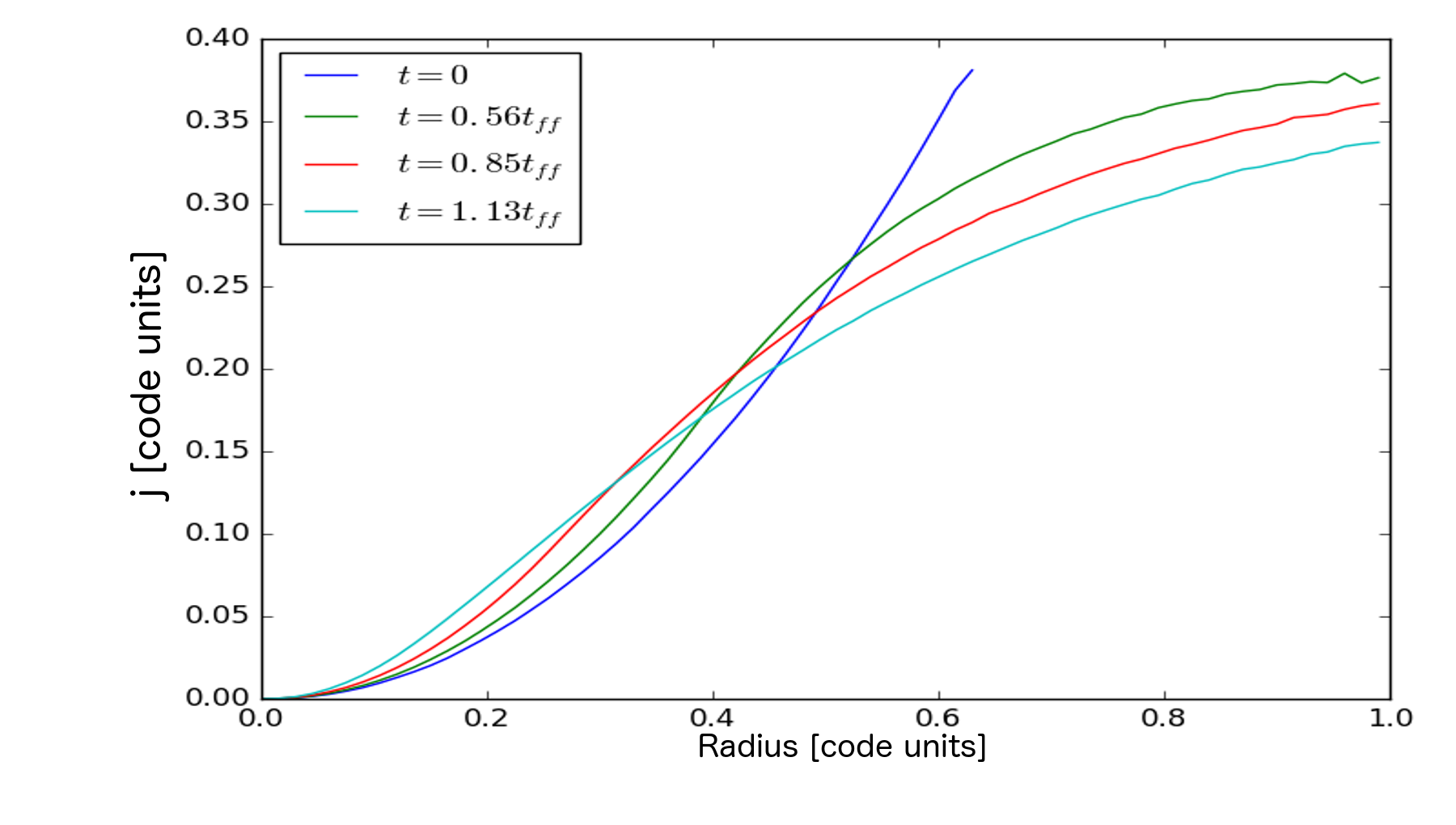}                                                                  
  \caption{The evolution of the specific angular momentum profile in the MFM collapsing rotating isothermal sphere run with 1M particles}\label{f:jprofile}                                                              
\end{figure*}

We run this setup with 10K, 100K, 1M particles, using both MFM and TSPH. All the simulations share the same properties discussed below.  In figure \ref{f:jprofile}, we plot the specific angular momentum profile of the sphere at different times.

Initially the sphere is uniformly rotating, and the special angular momentum $j\propto r^{2}$, where $r$ is the radius. The initial cloud radius is about 0.6 code length units. When material starts to flow inwards, angular momentum must be transported outward in order to conserve the total angular momentum of the system. Particles at the outer part of the cloud drift further. We show  the angular momentum profile out to only one code length unit because the outer part is noisy due to low number of particles.

\begin{figure*}[ht!]                                                                      
  \plotone{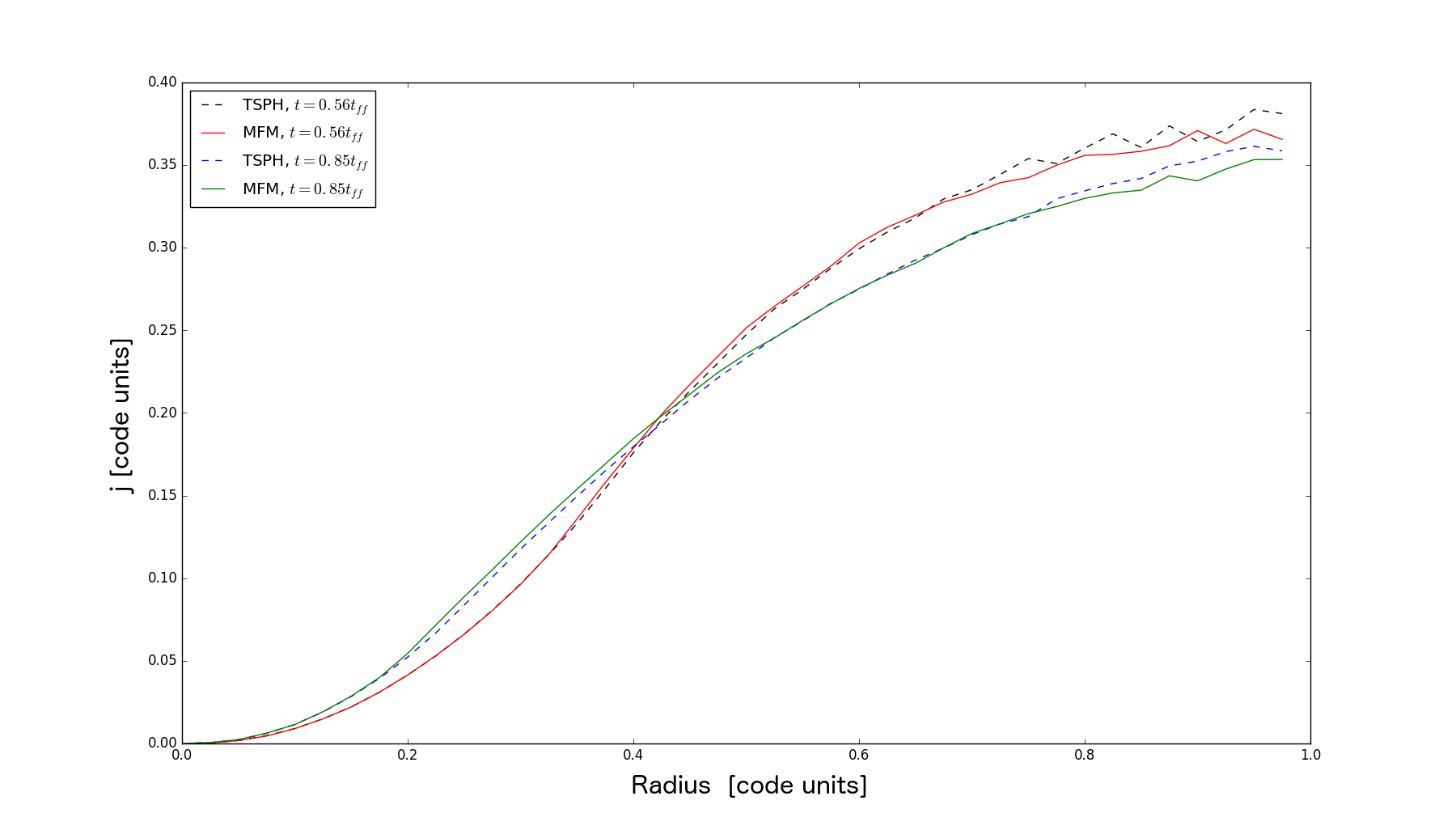}                                                                    
  \caption{Specific angular momentum profile with both MFM and TSPH, using 100K particles. At fixed time, the angular momentum profile evolves slightly faster in TSPH due to more significant numerical transport.}\label{f:hydro}                                                              
\end{figure*}

\begin{figure*}[ht!]                                                                      
  \plotone{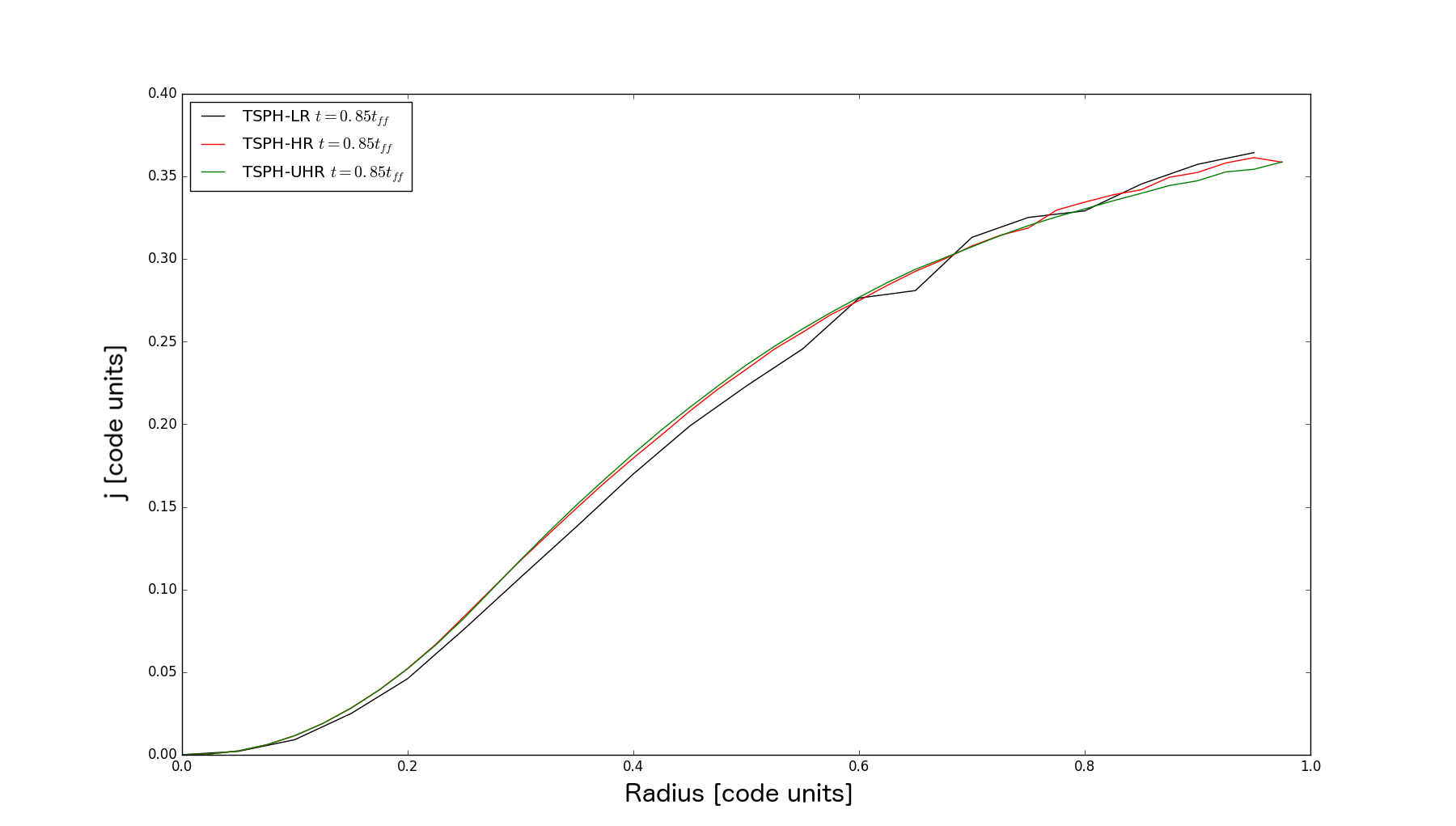}                                                                    
  \caption{Resolution dependence of the angular momentum profile in TSPH runs, using 10K (LR), 100K (HR) and 1 M particles (UHR)}\label{f:resol1}         
\end{figure*}

The angular momentum profile evolves less at higher resolution in \ref{f:resol1}, which is a sign of less numerical dissipation. The is because of the nice continuous flow property and numerical viscosity scales with $\it h$. 
From figure \ref{f:hydro}, we see very similar evolution of angular momentum profile by MFM and TSPH. However, at fixed time, the TSPH angular momentum profile(dashed lines) 
has evolved a bit faster, ie, more angular momentum has
been transported outwards, which is reflected by the 
fact that the MFM curve is above the TSPH curve inside 0.6 code length units and below outside.

We caution that the changing nature of the flow with increasing resolution,which we have shown to be a
central aspect of unstable disk evolution, does not play
a role here. This also means there are no sites of strong
shear and correspondingly strong numerical dissipation as
in the disk simulations. Therefore the fact that MFM
appears to perform only marginally better than TSPH is
not surprising as this is an easy flow configuration to
handle.

\section{Accreting planet test}
\label{sec:test2}
We are interested in assessing how numerical viscosity affects the angular moment transport, especially in cusps at interface
regions where high velocity derivatives occur.
. To avoid the complexity of resolving a full disk but carry out a test that can address a configuration with  velocity derivatives, we use a 2D shearing sheet \citep{Hawley1995} with an active accreting planet in the center to mimic the collapse under self-gravity. The simulations are similar to those in \citet{Ormel2015a} while the gravity of the planet is added following \citet{Ormel2015}.  We choose the local sound speed and the disk scale height as the natural units of speed and length, respectively. We set the gravitational constant to 1. The planet’s dimensionless mass $m \equiv GMp/(c^{3}_{s}/\Omega)$ equals its Bondi radius in dimensionless units $ R_{Bondi}/H$. In these units the dimensionless Hill sphere $r_{Hill} = R_{Hill}/H = (m/3)^{1/3}$.

The gravity of the planet is added $   F(t)=F_{0}\{ 1-exp[-\frac{1}{2}(\frac{t}{t_{inj}})^{2}]\}$. The force increases gradually to mimic the collapse due to self-gravity. $t_{inj}$ is the injection time. The gravity from the planet is smoothed to avoid numerical instability and increase the efficiency of calculation. $F_{0}=-\frac{m}{r^{2}}exp[-A(\frac{r_{in}}{r})^{p}-B(\frac{r}{r_{out}})^{q}]$. We set A=10, p=8 and B=1, q=4. The force transits from a pure Newtonian force $F_{0}=-m/r^{2}$ for $r_{in}<r<r_{out}$ to zero continuously and quickly. We set $r_{in}=0.1, r_{out}=m$ to prevent boundary condition corruption by the gravity of the planet.

\begin{figure*}[ht!]                                                                      
\plottwo{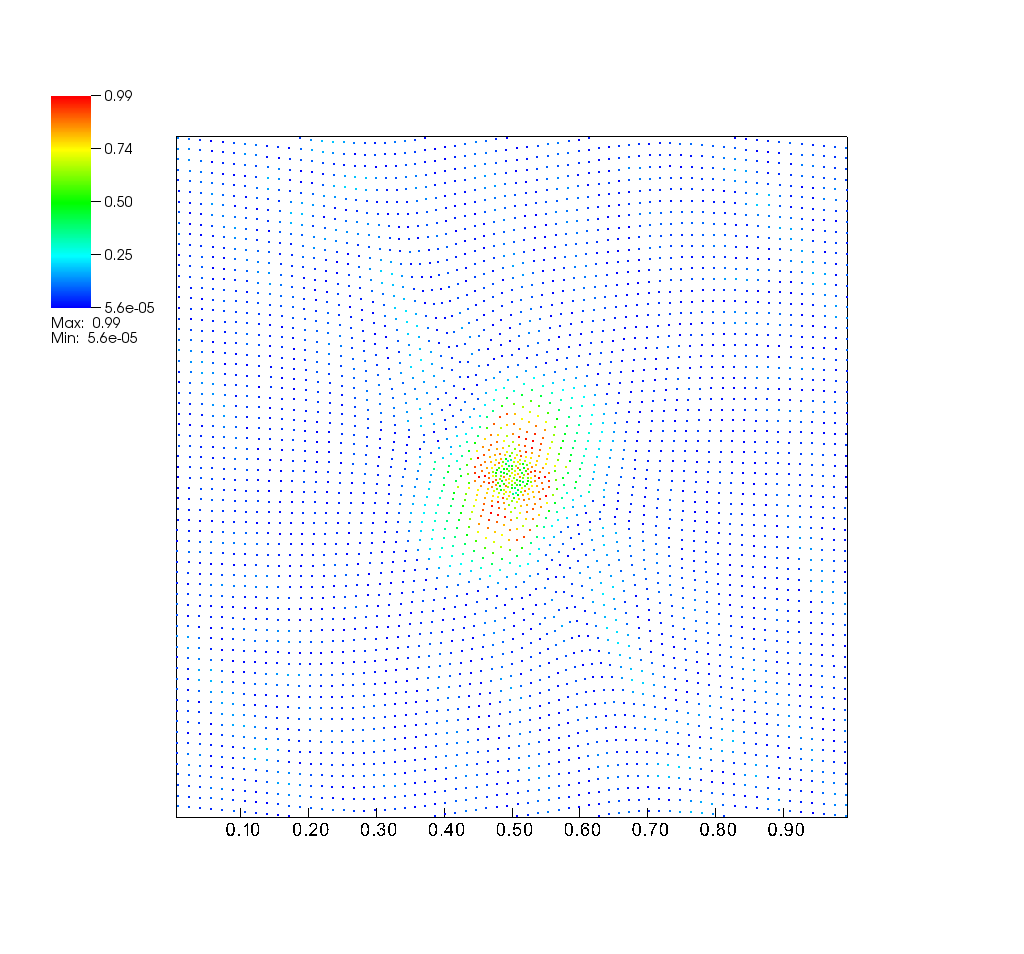}{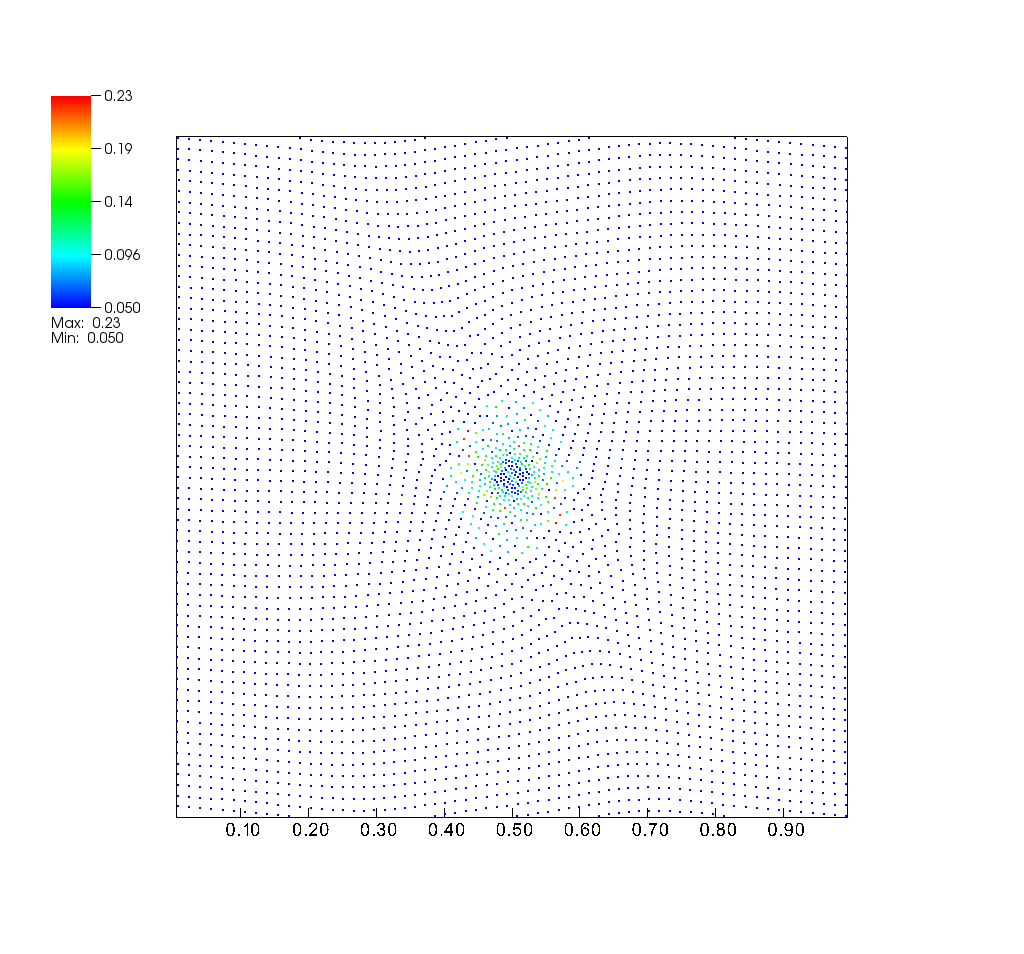}                                                          
\caption{Artificial viscosity $\alpha$ coefficient in TSPH (left) and ISPH accreting sink test (right) at $t=2$. ISPH has smaller artificial viscosity coefficient around the accretion flow due to the Cullen and Dehnen switch.}\label{f:sphav}                                            
\end{figure*}

\begin{figure*}[ht!]                                                                      
  \plotone{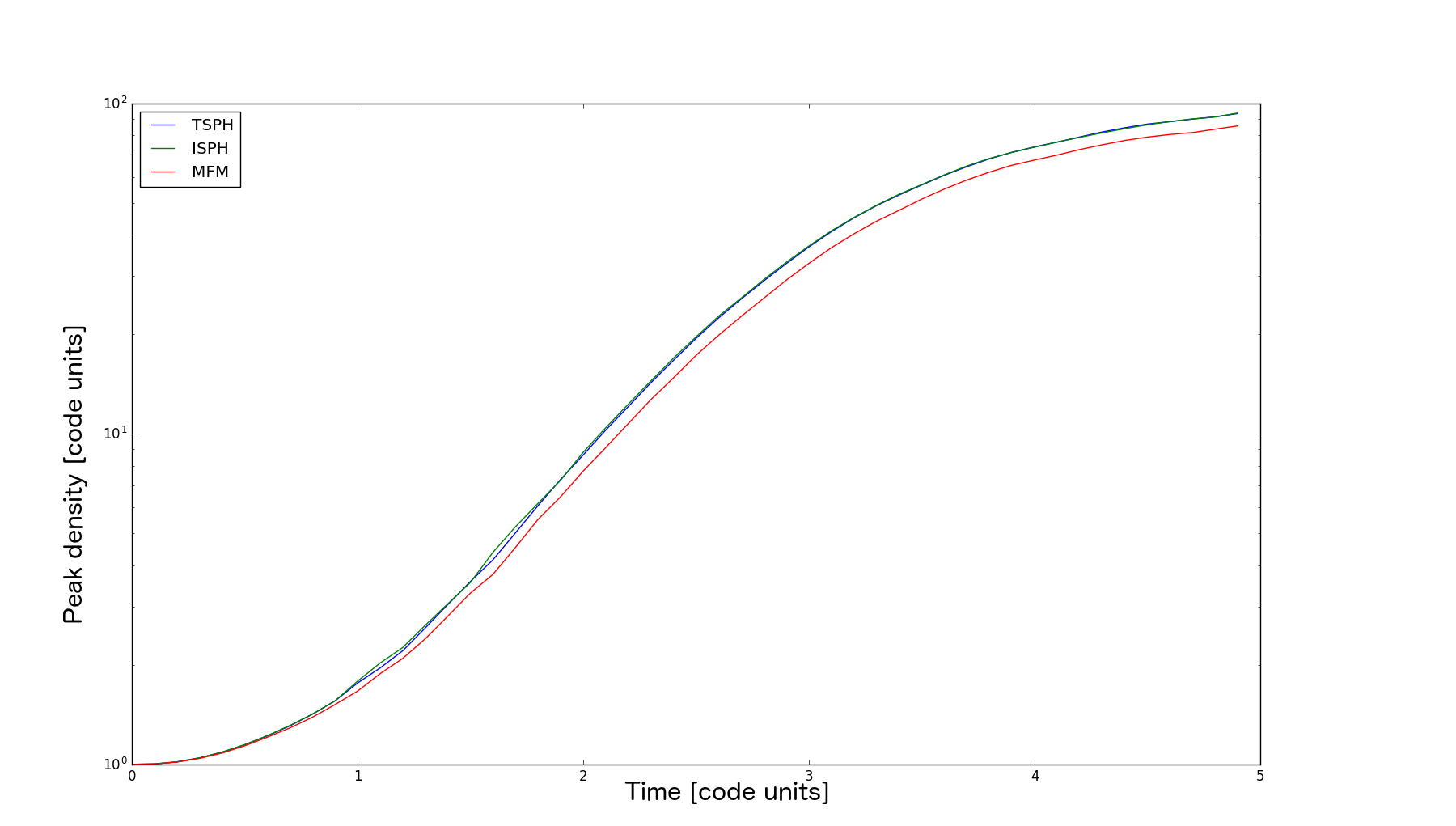}                        
  \caption{The peak volume density evolution with SPH and MFM. MFM always has the hghest peak density, reflecting
  a lower accretion rate.}\label{f:pkdt}                                                   
\end{figure*}

 We run simulations with $m=0.2, t_{inj}=2$ at a resolution of $64\times 64$ particles (this means 64 particles per scale height, which is even higher
 resolution than the 16 million particles UHR 3D disk runs used in this  paper).  The artificial viscosity $\alpha$ coefficient is shown in figure\ref{f:sphav} to compare
 directly TSPH and ISPH. ISPH, as expected, has
 lower artificial viscosity.
We use the time evolution of peak density to measure 
 how much mass in accreted over time inside the planet radius $r_{in}$as shown in figure \ref{f:pkdt}. The
 faster the growth of peak density the faster must be the
 outward transport of angular momentum resulting from
 any form of numerical dissipation, including, but not
 only, explicit artificial viscosity.
 Figure \ref{f:pkdt} highlights the lower dissipation of MFM.
 Indeed  MFM always has the lowest peak density while ISPH and TSPH are almost identical despite the lower value
 of the $\alpha$ coefficient.
 We argue the faster accretion in SPH methods is due to higher numerical viscous dissipation. The Cullen \& Dehnen switch is most effective away from shocks and it is not able to effectively suppress numerical viscosity in a
 flow with a large velocity gradient as in this test.

\end{document}